\magnification=\magstep1
\tolerance 500
\rightline{IASSNS-HEP-96/59}
\vskip 3 true cm
\centerline{{\bf Time and the Evolution of States in Relativistic
Classical and Quantum Mechanics}\footnote{*}{To be presented in a talk at the
conference on Physical Interpretations of Relativity Theory, Imperial
College, 6-9 September, 1996}}
\vskip 1 true cm
\centerline{Lawrence P. Horwitz\footnote{**}{On sabbatical leave from
 the School of Physics and Astronomy, Raymond and Beverly Sackler
Faculty of Exact Sciences, Tel Aviv University, Ramat Aviv, Israel,
and Department of Physics, Bar Ilan University, Ramat Gan, Israel.}}
\bigskip
\centerline{School of Natural Sciences, Institute for Advanced Study}
\centerline{Princeton, N.J. 08540}
\vskip 2 true cm
\noindent {\it Abstract\/}: A consistent classical and quantum
relativistic mechanics can be constructed if Einstein's  covariant
time is considered as a dynamical variable.  The evolution of a system
is then parametrized by a universal invariant identified with Newton's
time.  This theory, originating in the work of Stueckelberg in 1941,
contains many questions of interpretation, reaching deeply into the
notions of time, localizability, and causality.  Some of the basic
ideas are discussed here, and as an example, the solution of the two body  
problem with invariant action-at-a-distance potential is given.   A proper
generalization of the Maxwell theory of electromagnetic interaction,
implied by the Stueckelberg-Schr\"odinger dynamical evolution equation
and its physical implications are also discussed. It is also shown
that a similar construction occurs, applying similar ideas, in the 
theory of gravitation.

\vfill
\eject
\noindent{\bf 1. Introduction}
\smallskip
\par One of the most profound and difficult problems of theoretical
physics in this century has been the construction of a simple,
well-defined theory which unites the ideas of quantum mechanics
and relativity.   In fact, such a problem has existed in the consistent
formulation of classical relativistic dynamics as well.  A central
problem in formulating such a theory
is posed by the description of the state of a system which has
{\it spatial} properties (we shall come later to a discussion of
spin). 
\par  Nonrelativistic quantum mechanics, making
explicit use of the Newtonian notion of a universal, absolute time,
provides a description of such a state in terms of a square
integrable function over spatial variables at a given moment
of time.  This function is supposed to develop dynamically,
from one moment of time to another, according to Schr\"odinger's
equation, with some model Hamiltonian  operator for the system.
\par This description of a state is inconsistent with special
relativity from both mathematical and physical points of view.
The wave function described
in a frame in motion with respect to the frame in which the state
is originally defined takes on the meaning of a probability amplitude
at a set of different times, depending on the value of the spatial
variables.
Since the Hilbert spaces associated with different times are
distinct,  it therefore loses its interpretation as the description
of a state.
\par The situation for classical nonrelativistic mechanics is
quite analogous; the state of a system is described by a set
of canonical coordinates and momenta (the variables of the phase
space) at a given time.  These canonical variables develop in
time according to the first order Hamilton equations of motion.
\par  The variables of the phase space, under
the transformations of special relativity, are mapped into a new set
in which the time
parameter for each of them depends on the spatial location of
the points; in addition, there is a
  structural lack of covariance
of the phase space variables themselves.  The restoration of the
original description by extrapolating the dynamical evolution
through this family of times is not satisfactory since, in addition
to being highly impractical, the
specification of a state should be independent of the model for the
dynamical evolution of a system.
\par On the other hand, observed interference phenomena, such as the
Davisson-Germer experiment, showing the interference pattern due to the
coherence of the wave function over the spatial variables at a given
time, clearly should remain when observed from a moving frame (detectors
in motion relative to the original experiment).  Hence one would
expect that there is a more general, covariant, description of the
state of a system, which would predict such an interference pattern,
modified only by the laws of special relativity.
\par In the following, I describe the fundamental problem of
localization introduced by the use of the usual relativistic wave
equations [1].  I then turn to a discussion of Einstein's notion of
the (covariant) time, and show that it is possible to construct a
manifestly covariant classical and quantum theory in a framework,
first proposed in a complete form (for a single particle) by
Stueckelberg [2] in 1941, and, much later, developed for systems with
more than one particle [3], in which localization is rigorous. This
framework involves treating all four components of energy-momentum as
independent variables, reflecting the understanding of the Einstein
time as a non-trivially measureable quantity, and does not
restrict a particle to an infinitely sharp mass shell.  One therefore
considers {\it reducible} representations of the Poincar\'e group.  Some
fundamental questions, such as achieving an understanding of the
Newton-Wigner operator [4], and the Landau-Peierls uncertainty
relation [5] are then discussed.  As a concrete application, I discuss
the solution of the two body problem. The extension of the theory to
$U(1)$ gauge invariant
form, introducing electromagnetic interaction [6], involves a proper
generalization of the Maxwell theory which raises interesting
questions for the structure of general relativity as well.  
 \bigskip
\noindent {\bf 2. The problem of localization for the solutions
of relativistic wave equations.}
\smallskip
\par Attempts to take into account the required relativistic covariance
of the quantum theory by means of relativistic wave equations such as
the Klein-Gordon equation for spin zero particles, and the Dirac
equation for spin $1/2$ particles, have not succeeded in resolving
the difficulties associated with the definition and evolution of
quantum states.  These equations are of manifestly covariant form,
with the potential interpretation of providing a description
of a quantum state, with spatial properties, in each frame,
evolving according to the time parameter associated with that frame.
Newton and Wigner [1], however, have shown that the solution $\phi(x)$,
for example, of the Klein-Gordon equation, cannot have the
interpretation of an amplitude for a local probability density.
The function $\phi_0(x)$, corresponding to a particle localized
at ${\bf x}=0$, at $t=0$, has support in a range of ${\bf x}$
of order $1/m$, where $m$ is the mass of the particle. They
found that the distribution corresponding to a localized particle
is a (generalized) eigenfunction of the operator
$$ {\bf x}_{NW} = i \bigl( {\partial \over \partial {\bf p}}
- {{\bf p} \over 2E^2} \bigr). \eqno(2.1)$$
We shall return to discuss this problem, and its solution,
 in more detail later.
A similar conclusion was found for the solutions of the spin $1/2$
Dirac equation. The well-known problem posed by the lack of
a positive definite probability density for the Klein-Gordon
equation is formally managed by passing to the second quantized
formalism; the Dirac equation admits a positive definite density,
but the problem of localization remains.  In both cases, in
the second quantized formalism, the vacuum to one particle
matrix element of the field operator, which should have a quantum
mechanical interpretation (the one-particle sector), poses the
same problem of localization. The prediction of the formation
of interference patterns remains ambiguous in this framework.
\par   Foldy-Wouthuysen [7] type transformations (for both spin zero
and spin $1/2 $ cases)
restore the local property of the wave functions [3], as for the
non-relativistic Schr\"odinger theory, but in this representation,
manifest Lorentz covariance is lost. It is clear that the problem of
localization is a fundamental difficulty in realizing a covariant
quantum theory by means of the usual wave equations.
\par Hegerfeldt [7] has shown, moreover, that a wave function
defined at time $t=0$ which is localized in the sense of
compact support, or in the Newton-Wigner sense, or has fast fall-off
in spatial directions, cannot maintain this degree of localization
at time $t>0$ if the free evolution is governed by such wave
equations (it necessarily evolves out of the light cone).
\vfill
\eject
\noindent {\bf 3 The Einstein notion of time.}
\smallskip
\par  The resolution of the problem of localization posed above
lies in the formulation of a quantum theory for particles which
are not precisely constrained to a pointwise mass-shell, i.e.,
the relativistic particle is not restricted, in its definition, to
an irreducible representation of the Poincar\'e group as advocated
by Wigner in his fundamental paper of 1939 [9].  The kinematic
definition
$$      E^2 = {\bf p}^2 + m^2  \eqno(3.1)$$  
is, of course, maintained, but the quantity $m^2$ is to be considered
as a {\it dynamical variable}, with values determined by the
interactions in the physical system [3]. If one wishes to construct a 
theory of particles,
on a phenomenological level, which does not make direct use of the
energy stored in the fields that surround them, it is necessary
to admit this degree of freedom.  For example, the electron is seen
to have an effective mass, in the context of Lamb shift
calculations [10], differing from its free value when it is bound in
an atom, and the effective masses of nucleons bound in the nucleus
differ from their free values. One sees immediately from the
assumption of this added degree of freedom that, in the quantum
theory, the operators corresponding to the mass and the position
of the particle are incompatible. In this way, one can understand the
foundation of the Newton-Wigner problem [3], as follows.
\par The scalar product of Klein-Gordon ``wave functions'' is given by
$$ \int {d^3{\bf p} \over 2E } \phi^*({\bf p}) \phi({\bf p}) =
(\phi_1, \phi_2). \eqno(3.2)$$
Newton and Wigner start by assuming that $\phi_0({\bf p})$ is the wave
function of a particle localized precisely at ${\bf x} = 0$; they
then assume that the function $\phi_a({\bf p}) = \exp {i{\bf p}\cdot
{\bf a}}$ is orthogonal to $\phi_0$ under the scalar product $(3.2)$.
The integral then corresponds to the Fourier transform of the function
$|\phi_0|^2({\bf p})/{2E}$; since it vanishes for all ${\bf a}$ not
zero, the integrand must be constant. This implies that 
$$ \phi_0({\bf p}) = C\sqrt{E}, \eqno(3.3)$$
and therefore that the Fourier transform, which is the wave function
in space, is not a $\delta-$function (as it would be if
$E=\rm{const}$, as in the nonrelativistic limit), but a function which
is spread out to the Compton wavelength of the particle.  Hence these
wave functions do not localize the particle.  Another way to undertand
this, in the framwork of the Stueckelberg theory, is that the variable
${\bf x}$ does not commute with $m^2$, as defined in $(3.1)$, which
explicitly contains the momentum variable, and hence the mass of a
particle and its position are incompatible. In the discussion of the 
covariant relativistic quantum theory below, I explain how the
Stueckelberg theory is indeed local in its description of the
elementary objects of the theory, points in spacetime that might be
called ``events,'' and in a decomposition of the expectation value of
the position ${\bf x}$ over all mass shells, one finds the
Newton-Wigner operator $(2.1)$ at each sharp mass value in the
integral.
\par If, as I have argued above, we consider the mass of a particle as
a dynamical variable,  ${\bf p}$ and
$E$ must be considered as independent dynamical variables.
The complementary variables, dual to these, are ${\bf x}$ and $t$,
and we must conclude that these are also independent
 dynamical variables. To understand the role of $t$ as a dynamical
variable, rather than a c-number parameter, we must examine
carefully the Einstein notion of time.
\par To do this, let us examine the structure of the Lorentz
transformation, restricting for now our discussion to one space
and one time dimension,
$$ \Delta t' = {{\Delta t - v\Delta x}\over  \sqrt{1-v^2}},
\eqno(3.4) $$
where we take $c$, the velocity of light in vacuum, to be unity.
  This expression, in Einstein's original
interpretation, corresponds to the difference between two values of
time recorded by detectors equipped with clocks in a frame $F'$
moving at velocity $v$ with respect to a frame $F$ in which the
two signals are emitted in an interval of time $\Delta t$, at two
places separated by an interval $\Delta x$.  One may think of
the two events as associated with the same emitting object; in this
case, it follows that the emitter has moved from $x_1$ to $x_2$
in the time $t_2 - t_1$.  If the emitter is at rest in $F$, then
both signals are sent from the same point, and Eq.$(3.4)$
reduces to
$$ \Delta t' = {{\Delta t} \over {\sqrt{1-v^2}}} \equiv {{\Delta s}
\over {\sqrt{1-v^2}}},       \eqno(3.5) $$
where $\Delta s $ is usually referred to as the proper time of the
emitter in $F$.
\par It should be emphasized that the difference between $\Delta t'$
and $\Delta s$ is not at all associated with the distance
between the detectors in $F'$ and the emitters in $F$.  We may
consider the detection as carried out by a set of receivers
in $F'$ filling, in the sense of Wheeler and Taylor [11], all of 
spacetime with synchronized clocks relatively at rest in $F'$,
and in uniform motion relative to the emitter in $F$.  The detectors
that lie in the immediate neighborhood of the emitter in $F$ at
$t_1$ and $t_2$ receive the signals with no retardation due to
 spatial separation; since all the detectors are synchronized,
the two readings obtained in this way, generally in two different
detectors of $F'$ (due to the uniform motion of this frame), $t_1'$
and $t_2'$, can be treated as being detected by equivalent clocks
in $F'$, i.e., a global detection in the $F'$ laboratory.
\par It is essential that the clocks associated with with the
detectors in $F'$ be precisely physically equivalent to the clocks
associated with the emitters in $F$, and set to run at the same rate,
so there can be a basis for the comparison between $\Delta t'$ and
$\Delta t$.  This assumption, that there exists a {\it standard
 clock} which can be found embedded in every frame independent
of its (uniform) motion, is implicit in any discussion of
special relativity.  It is, furthermore, an essential postulate
of general relativity as well.
\par As we know from general relativity, the apparent rate of a
clock in a gravitational field is affected by a change in the
metric tensor (the mechanism of the gravitational redshift).
Local inertial frames, according to the equivalence principle,
are characterized by motion in which they are freely falling
(moving along a geodesic), so that the clocks imbedded in these
frames are not subject to the influence of the gravitational
field.  This condition must be met for the inertial frames
$F$ and $F'$ that we discussed earlier.  Moreover, if the clocks that
we consider have a varying self-energy caused by springs under tension
or batteries with stored chemical energy, the rate of recording
time of these clocks may be affected by the gravitational fields
induced by the corresponding local concentration of energy density.
The standard universal clocks that we visualize as imbedded in each
inertial frame must therefore be {\it ideal} clocks, in the sense that
they contain no self-energy induced frequency shifts.  In this
case the proper time interval referred to in Eq.$ (3.5)$ recorded by an
ideal detector at rest with respect to the emitter is $\delta s = \Delta \tau$,
where we have designated the interval on the ideal standard inertial
clocks as $\Delta \tau$.
 \par The notion of time, as generated by an ideal standard inertial
clock existing, in principle, in every inertial frame (locally, in
free fall, in the presence of gravitational fields), coincides
precisely with the universal time postulated by Newton [12], and in
terms of which equations of motion for systems in interaction may be
formulated. The formulation of a theory of mechanics with ``action at
a distance'' interaction, of the form $V(x^\mu_1, x^\mu_2)$ would, in
fact, be meaningless without the postulate of such a universal time,
since the choice of $x^\mu_1$ and $x^\mu_2$ as two points along the
respective world lines must be determined by some agreement about
which two points are in interaction.  This agreement, or correlation,
is determined by the assumption of the existence of a universal
parameter $\tau$.  I do not claim that it is possible to synchronize 
clocks in relative motion.  The ideal clocks in any frame may be set
to any arbitrary initial value. What we explicitly assume, as a
postulate, is that {\it there exists a universal time by means of
which dynamical interactions are correlated}.
\par The assumption of a single universal time $\tau$ of the world,as
for the Galilean time in nonrelativistic dynamics, has far-reaching
consequences.  An interesting question, closely related to the twin
paradox, immediately arises in general relativity.  We have
operationally defined the universal time parameter $\tau$  
 in correspondence with the reading on an ideal inertial clock.
However, the integral of $d\tau = \sqrt{-g_{\mu \nu}dx^\mu dx_\nu}$
[we use the metric $(-,\,+,\,+,\,+)$ in the  local flat space limit]
along two different geodesics which start at a common point and later
cross at another point in spacetime, give two different results, in general.
 Hence, one may argue that this operational definition is not
consistent with a universal $\tau$.  This argument rests on the
generally accepted idea that two points which coincide in (local)
Minkowski  space correspond to the same object [13].  It arises  from
our use of Minkowski space as a complete coordinatization of physical
phenomena.  We are, however, free to postulate that two events which
overlap in spacetime, but have different values of the universal time
$\tau$,  are {\it not} identically the same physical objects.  The
separation can be easily seen
graphically by adding $\tau$ as an additional coordinate in
configuration space.  One then sees that the projection of the
geodesics onto spacetime appear to cross when the two trajectories do
not, in fact, coincide.  Whether the two objects at $\tau_1,\,
\tau_2$ over this intersection can be in nontrivial interaction
depends upon assumptions about the form of the interaction (and the
properties of the appropriate Green's functions [14]).  In this picture
the ``twins'' age in free fall in precisely the same way, by
definition. The potential model we considered earlier corresponds to
interaction only at equal $\tau$. The twins are never faced with a
paradox.
\par As an example of the type of interactions that may take place
in the neighborhood of a locally flat region, the propagators for
generalized (as we shall discuss later) electromagnetic fields in 
this framework may have spacelike components
for domains in Minkowski space which are separated by distances
greater than the $\tau$ difference [16].  For dynamical correlations
to exist it is therefore necessary that the $\tau$ difference be less
than the scale of the spacelike separation; for nearby  spacelike
neighbors, such interaction therefore requires almost equal $\tau$
origin of signal and dynamical impact, such as in the potential models
referred to above.  I do not use the term transmission of
``information'' here, since the classical notion of information, as
we shall see below, requires integration over $\tau$; in this case the
spacelike components of the Green's function vanish.
\par To conclude this section, let us consider the simple example of
the gravitational redshift and its dynamical implications. Suppose
that an ammonia molecule on the planet Jupiter radiates (we adopt
here, for the sake of
illustration, a semiclassical picture) due to the motion of the
nitrogen atom through the plane of the three hydrogen atoms. The
interval $\Delta t_J$ between oscillations is related to an inertial
proper time interval $\Delta \tau_J$ by the generic formula 
$$ \Delta t' = {\Delta \tau \over \sqrt{-g_{00}}}, \eqno(3.6)$$
 obtained from the equation
$$ d\tau^2 = -g_{\mu \nu} dx^\mu dx^\nu \eqno(3.7) $$
for the invariant interval of general relativity.  Here, the
differential $dx^\mu$ for a clock at rest on the surface of Jupiter is
taken to be just $dt_J$ , so that only the $g_{00}$ term of the metric
tensor $g_{\mu \nu}$ appears. Applying the same formula to the
intervals on Earth, one then divides one by the other (for example,
ref. [15]), to obtain
$$ {\Delta t_J \over \Delta t_E} = \sqrt{g_{00}^E \over g_{00}^J }
{\Delta \tau_J \over \Delta \tau_E}; \eqno(3.7)$$
 assuming that ${\Delta \tau_J \over \Delta \tau_E}$ is unity, one
obtains a results that agrees very well with experiment.  The
experiment raises two important basic questions: a) By what physical
mechanism can the ammonia atom control its oscillation frequency
relative to the time of a freely falling clock? b) Why should the
proper time interval associated with the oscillation of two clocks,
one on Earth and the other on Jupiter, be the same, so that
cancellation can take place? The ammonia atoms are not in free fall,
bound to the surface of their respective planets, and there is no
obvious mechanism for the radiation frequency to be related to the
time of a freely falling clock.  Furthermore, whatever mechanics
governs the motion of the atom, it is not clear that the proper time
interval for emission of one cycle of a wave should be the same on
Jupiter and the Earth, 
\par These questions can be simply answered by assuming that there is
a universal world time which drives the motion of the ammonia clocks
in both environments, and that the radiation which is emitted is
affected by the gravitational field.  The differential equations
satisfied by the atomic motions would contain the independent variable
$\tau$ corresponding to this universal evolution, just as Newton's
equations contain the Newtonian time in the nonrelativistic theory of
mechanics.  In fact, one may postulate that all machines run according
to this universal time.  The freely falling clock of Einstein is also
a machine, and since there are no forces acting on it, and it is ideal
(no batteries, springs which can wind down, or any other sources of
change in its self-energy), it may be set to read the universal time
on its face. The two ammonia clocks therefore run by the same
universal time, and the fact that these configurations are governed by
oscillations according to this rate, the radiation is emitted at this
basic frequency (governed by the mechanical system), and modified by
the acceleration intrinsic to the local gravitational field according
to the redshift formula.
\par This example illustrates the necessity of postulating a universal
time which is not a clock, but a time according to which the clocks
and all other machines run.  In fact, it is logically not satisfactory
to define time according to clocks, for the simple reason that clocks
must run according to the flow of time themselves.
\bigskip
\noindent{\bf 4. Covariant Classical and Quantum Mechanics}
 \smallskip
\par Stueckelberg's original paper discussed the possibility of pair
annihilation and production in {\it classical} relativistic mechanics.
He first considered the motion of an event in spacetime tracing out a
free worldline, clearly a straight line moving within the light
cone, as for the kinematics of a freely moving massive particle.  He
then points out that interaction can bend this worldline, as might be
generally expected, and then moves on to the possibility that the
worldline can bend so much that it curves back in time, that is, after
reaching a maximum position along time axis, it turns and heads
backward in time.  The interpetation of such a collection of events in
the laboratory is that of two objects moving toward each other in the
sequence of time signals generated in the detectors of the laboratory
for the occurrence of events (e.g., with Geiger counters, or in a
spark chamber) until they meet and annihilate.  This is the classical
possibility of pair ahhihilation.  Stueckelberg then points out that
the time $t$ is no longer adequate, under these conditions, to serve as
an independent variable, since the curve would then not be a function
(two values of the space variable correspond to one of the time).  He
introduced a parameter along the trajectory, $\tau$, which then serves
as the effective independent variable.  Due to the reversal of four
momentum in space-time, one easily sees that the object moving backward
in $t$ would appear to have opposite charge if it were thought of as
having positive energy, and going {\it forward} in $t$;  it is
therefore identified as the antiparticle.  This interpretation of the
antiparticle  was used
by Feynman [17] in his construction of the standard quantum
electrodynamics from a perturbative point of view in 1949. 
\par Stueckelberg then proceeded to construct a classical mechanics in
spacetime by defining a function $K$ on the phase space which
satisfies the Hamilton equations
$$ {dx^\mu \over d\tau} = {\partial K \over \partial p_\mu} \qquad
{dp^\mu \over d\tau} = {\partial K \over \partial x_\mu}, \eqno(4.1)$$
    where the index $\mu$ takes on values $0,\,1,\,2,\,3$
corresponding to the physical coordinates of time and space.   These 
relations can be derived
 from a variety of variational principles, as for the usual
nonrelativistic mechanics, extended to four dimensions. For the free
particle, one may choose
$$K_0 = {p^\mu p_\mu \over 2M}, \eqno(4.2)$$
where M is a dimensional scale parameter, an intrinsic property of
the particle.  The equations of  motion then become
$$ {dx^\mu \over d\tau} = {p^\mu \over M}, \,\, {dp^\mu \over d\tau}
= 0 \eqno(4.3)$$
so that, dividing the space components by the time component in the
first of these, one finds
$$ {d{\bf x} \over dt} = {{\bf p} \over E }, \eqno(4.4)$$
where $p^0 \equiv E$, precisely the Einstein relation for velocity
induced by setting the observational frame into motion with velocity
$-{\bf v}$ with respect to the particle.  Note that from $(4.3)$,
$$ \eqalign{{dx^\mu \over d\tau}{dx_\mu \over d\tau} &= {p^\mu p_\mu
\over M^2} \cr &=-\bigl({ds \over d\tau}\bigr)^2, \cr}
\eqno(4.5)$$
from which we see that the proper time squared $(ds)^2 = -dx^\mu
dx_\mu $ is equal to $(d\tau)^2$ only if $p^\mu p_\mu = -M^2$, the
so-called ``mass shell'' condition. In case there are batteries or
springs that can run down, $K$, which is $K_0 + V$, may still be
constant, and then $K_0$ changes with the changing $V$. In 
the nonrelativistic limit of
the theory, one finds that $M$ may play the role of the Galilean
target mass [18].  Piron and I [3]
extended this construction to systems of arbitrary $N$ events; the
corresponding $N$ worldlines correspond to $N$- particle systems.  The
generalized Hamiltonian for such a system might be chosen, for example
as the many-body action-at-a-distance form
$$ K = \sum_i {p^\mu_i p_{\mu_i} \over 2M_i} \,\, +\,\,V(x_1,
\dots,x_N). \eqno(4.6)$$
 Stueckelberg proceeded to construct the quantum theory associated
with this symplectic mechanics; one assumes the covariant commutation
relations 
$$ [x^\mu,\, p^\nu] = i g^{\mu\nu}, \eqno(4.7)$$
and the (Stueckelberg-Schr\"odinger) equation
$$ i{\partial \psi_\tau (x) \over \partial \tau } = K \psi_\tau (x),
\eqno(4.8)$$
a Shr\"odinger-like equation first order in $\tau$.  Here, we have
denoted by $x$ the four dynamical variables $x^\mu$.  The wave
function $\psi_\tau(x)$ belongs to a Hilbert space over $R^4$, and the
wave function then falls off integrably in both space and time directions.
The Fourier transform to energy-momentum representation,
$$ \psi_\tau (p) = \int d^4x e^{ip^\mu x_\mu} \psi_\tau(x),
\eqno(4.8)$$
enables us to express the evolution of a free particle as 
$$ \psi_\tau(x) = \int d^4p \exp\{ -i{p^\mu p_\mu \over 2M} \} e^{ip^\mu
 x_\mu} \psi_0(p), \eqno(4.9)$$
so that the stationary phase point moves, as in the standard Ehernfest
relation, approximately with the relations $(4.3)$, the classical
motion of a spacetime event along the free particle world-line.
\par In the nonrelativistic Schro\"dinger theory, one may demand that
the theory remain invariant in form under a $U(1)$ phase
transformation, of the form $\psi \rightarrow e^{i\Lambda}\psi $.  In the
representation of $(4.7)$ for which $p^\mu \rightarrow -i{\partial/\partial
x_\mu}$,  we see that in the equation $(4.8)$, $p^\mu$ must be
replaced by the covariant derivative $p^\mu-e_0a^\mu$ and , morevover,
the $\tau$-derivative, $i\partial_ \tau$ (where we use an abbreviated
form for the derivative) must be replaced by $i\partial_\tau +
e_0a_5$. Each of the gauge compensation fields $a_\alpha$, functions
of both $x$ and $\tau$, then undergo a gauge transformation [6].
$$  a_\alpha \rightarrow a_\alpha + {1 \over e_0} \partial_\alpha
\Lambda, \eqno(4.10)$$
thus restoring the equations to their original form.
\par To define dynamical equations governing the
structure of the corresponding  gauge invariant ``field
strengths''
$$ f_{\alpha \beta} = \partial_\alpha a_\beta - \partial_\beta a_\alpha,
\eqno(4.11)$$
we define  the Lagrangian density
$$ {\cal L} = {1 \over 4} f_{\alpha \beta} f^{\alpha \beta},
\eqno(4.12)$$
along with the Lagrangian density generating the
Stueckelberg-Schr\"odinger equation as a field equation.  The usual
variational principle results in the additional  field equation
$$ \partial_\beta f^{\alpha \beta} = j^\alpha, \eqno(4.13)$$
where $j^\alpha = \{j^\mu, \rho \}$ is the conserved {\it
five}-current 
associated with the Stueckelberg-Schr\"odinger
equation, derived in exactly the same way as the conserved current of
the nonrelativistic Schr\"odinger equation; $\rho$ is just
$|\psi_\tau(x)|^2$, and $j^\mu$ is the antisymmetric bilinear form for
$-i\partial^\mu - e_0 a_\mu$.  The conservation of $j^\alpha$ is consistent
with the antisymmetry of the field strengths $f^{\alpha \beta}$.  
\par It is a remarkable fact that, as we have seen from the Schr\"odinger
case in three dimensions and  the Stueckelberg case in four, that the
requirement of gauge invariance implies supplementing the manifold of
the measure space by one additional dimension.  A theory in which a
``world'' is described at a given moment of the evolution parameter is
supplanted by a world in which the evolution parameter is
incorporated, in the same second order as for the spatial (or
spacetime) variables, providing the homogeneity that could support
a higher symmetry. One may think of these gauge fields as correlation
fucntions between the sequence of  worlds parametrized on the
foliation provided by the evolution. 
\par  This apparent higher symmetry is not really
present when there is interaction with the matter fields, since the
density $\rho$ cannot transform with a current that contains derivatives,
and, as with the Schr\"odinger Galilean theory, they break this symmetry
locally. In this latter case, the symmetry of the free field
equations for the homogeneous, noninteracting, field equations is
O(3,1). Experimentally, it was found that this symmetry, the Lorentz
symmetry of the world, is, in fact, more accurate than the previously
assumed Galilean symmetry.  The fields associated with the  gauge
invariant extension of the Stueckelberg theory appear to exhibit an
O(3,2) or O(4,1) symmetry, depending on the choice of metric for the
raising and lowering of the fifth ($\tau$) index [6]. In the gauge for
which $\partial_\alpha a^\alpha =0 $, analogous to the Lorentz gauge, the
equations $(4.13)$ become, in the absence of sources,
$$ \partial_\alpha \partial^\alpha a_\beta = \pm \partial_\tau^2
-\partial_t^2 + \nabla^2 =0. \eqno(4.13)$$
Under Fourier transform with repect to $\tau$, the first term becomes
$\mp m^2$, where $m$ corresponds to the mass associated with this
component of the field.  The O(3,2) choice of metric therefore
corresponds to a field with non-physical mass (tachyonic), and may be
thought of as intrinsic radiation; its limit onto the light cone, for
$m \rightarrow 0$ corresponds to Maxwell radiation. The choice of the
O(4,1) metric carries a real physical mass, and could be used to
represent a physical vector meson, for example.   The dynamical
transiton from the O(3,2) to the O(4,1) metric may represent vector
meson photoproduction.
\par Note that the usual Maxwell theory is properly contained in this
generalized form, which we call pre-Maxwell.  Eq. $(4.13)$, for the
$\{\mu \}$ components alone, may be written as 
$$ \partial_\nu f^{\mu \nu} + \partial_\tau f^{\mu 5} = j^\mu .
\eqno(4.14)$$
Integrating over $\tau$ from $-\infty$ to $\infty$, with appropriate
asymptotic conditions, the second term vanishes.  The
current consevation law for the Stueckelberg current, 
$$ \partial_\alpha j^\alpha = 0, \eqno(4.15)$$
under similar integration, becomes
$$ \partial_\mu \int d\tau j^\mu = 0, \eqno(4.16)$$
a result known to Stueckelberg [2].   Hence, one sees, with
Stueckelberg, that $\int d\tau j^\mu$ is the Maxxwell four-conserved
current, and occurs on the right hand side of the integrated version
of $(4.14)$.  The
$\tau$-integral of the left hand side may therefore be identified
with the Maxwell fields $F^{\mu \nu}$; these {\it zero modes} of the
pre-Maxwell theory are just the Maxwell fields. The transmission of
information, according to antennas and receivers as we know them, is
associated primarily with the zero modes of the fields and currents.
\par   Finally, since $\int
d\tau a^\mu $ are the Maxwell vector potentials $A^\mu$, we see
that the dimension of $a^\mu$ is $ 1/L^2$, so that the ``charge''
$e_0$ must have dimension $L$.  The Lagrangian density $(4.12)$
requires an additional factor, say, $\lambda$, of dimension length.
Moreover, the field equations $(4.13)$ contain a current on the right
hand side which is, in reality, proportional to $e_0$.  The ratio
$e_0 / \lambda$ then corresponds to the Maxwell charge $e$, and
one understands the restriction to the neighborhood of the zero mode
as a property of statistical correlations, in $\tau$, of the field
(last of ref. [6]).
It has been shown that, in a path integral quantization of the corresponding
field theory, going to the limit in which only a small neighborhood of
the zero mode survives is equivalent to Pauli-Villars
regularization[19].
\vfill
\eject
\bigskip
\noindent{\bf 5. Applications}
\smallskip
\par In this section, I conclude by describing some applications of
the theory outlined above to give concrete illustrations of how these
constructions, based on the fundamental premise that the Einstein time
$t$ is a dynamical variable, may be manifested in known theoretical
frameworks as well as in the world represented by laboratory
experiments.
\par I first discuss the two important theoretical results of Newton
and Wigner [4] and of Landau and Peierls [5] mentioned above. Startin
witht he wave functions in the momentum representation defined in
$(4.8)$, we may ask for the quantum mechanical expectation value of 
$$ ({\bf x}_{op})|_{t=0} = i {\partial \over \partial{\bf p}} - {1 \over 2}\{t,\, {{\bf
p} \over E} \}. \eqno(5.1)$$
The second term is the shift $vt$ extrapolating an event occurring at
$(t,{\bf x})$ back to the zero-time axis.  The operator sought by
Newton and Wigner [4] was the one for which the spectrum provides the
values of ${\bf x}$ at $t=0$. In the computation of the expectation
value, 
$$ <{\bf x}_{op}> = \int d^4p \, \psi^*(p) {\bf x}_{op} \psi(p),
\eqno(5.2)$$
we may change variables using the definition $ E = \sqrt{{\bf p}^2 +
m^2}$, where $m$ is a variable.  Then, $d^4p = d^3{\bf p} dm^2 / 2E$.
For each value of $m^2$, the measure then coincides with that of Eq.
$(3.2)$.  However, the momentum derivative acting on the wave
function, intended to differentiate only the first three arguments,
would now act freely on all four.  We must therefore subtract the
contribution $(\partial E/\partial {\bf p})(\partial_E) = ({\bf
p}/E)(\partial_E)$.  This cancels precisely the contribution of this
term from the anticommutator, leaving, on each mass shell, the
operator $(2.1)$.  
 \par Landau and Peierls [5] studied the problem of the existence of a
measurement of the first kind on the spectrum of an operator with
continuous spectrum.  After a carefull heuristic analysis making use
of the time-energy uncertainty relation, they arrived at the
inequality
$$ \Delta p \Delta t \geq \hbar/c \eqno(5.3)$$
as an estimate.  There was some criticism of this relation, since it
was not derived in the same rigorous way that the position-momentum
relation is obtained as a Schwartz inequality in the quantum theory.
However, we may think of a Geiger counter which  measures the time of
occurrence of the passage of a particle at a point $x_0$, and suppose
that some event occurs at a time $t$ and place $x$ different from $x_0$. 
If that event is part of a world line, then it may be extrapolated
back by dividing the distance by the velocity, i.e., one can define in
analogy to the Newton-Wigner operator of Eq. $(5.1)$, the operator
$$ (t_{op})|_{x=x_0} = -i\partial_E - {1 \over 2} \{x-x_0, {E \over
p}\}. \eqno(5.4)$$
A calculation similar to that of the Newton-Wigner case discussed
above provides a formula for the effective time operator on each mass shell.
The commutator of $p$ with the operator $t_{op}$ does not vanish due
to the presence of the second term, and one finds precisely the
commutator necessary to achieve the Landau-Peierls relation $(5.3)$
rigorously, through the Schwartz inequality.
\par The crucial point in Hegerfeldt's proof [8] that acausal behavior
is associated with the Dirac and Klein-Gordon fields is that the
analyticity implied by the Fourier transform of compactly supported
functions is destroyed by the application of the evolution operator
$\exp\{-i\sqrt{{\bf p}^2 + m^2}\}$.  The operator $K_0$ is simply
quadratic in momentum, and hence analytic, so it may not change the
analytic character of compactly supported functions.  However, one
must take care that a boundedness requirement on the range of $p_\mu
p^\mu$ does not change this property; it appears that tachyon
components, or a balance between particles and antiparticles (with
some implications for the theory of detection) may be required [20]. 
\par A very important application of the two-body form of the theory
in illustrating the idea of the existence of a dynamical time variable
is that of the relativistic two body bound state.  As a model for an
action-at-a-distance potential, let us use the form $V(\rho)$, where
$$ \rho = \sqrt{ (x_1-x_2)^2 - (t_1 - t_2)^2 }, \eqno(5.5)$$
assuming that the two events described dynamically by the theory are
spacelike separated so that the square root is well defined.
Separating variables in the equation $(4.8)$ to total momentum $P^\mu
= p_1^\mu + p_2^\mu$ and relative momentum $p^\mu = (M_2 p_1^\mu - M_1
p_2^\mu)/ (M_1+ M_2)$, Eq. $(4.8)$ can be written
$$ i {\partial \psi_\tau \over \partial \tau} = {P^\mu P_\mu \over 2M} +
{p^\mu p_\mu \over 2m} + V(\rho), \eqno(5.6)$$
where $M= M_1 + M_2$ and $m$ is the reduced mass $M_1M_2 / (M_1 +
M_2)$. The center of mass motion represented by the first term may be
factored out by applying a phase to the wave function, and the
remaining reduced motion problem solved as a stationary eigenvalue
problem.  If the support of the wave function is taken as the full
spacelike region, one separates variables in the d'Alembertian
according to the definitions
$$\eqalign{ t&= \rho \sinh \beta \cr
            x_1 &= \rho \cosh \beta \cos \theta \cos \phi \cr
            x_2 &= \rho \cosh \beta \cos \theta \sin \phi \cr
            x_3 &= \rho \cosh \beta \sin \theta .\cr } \eqno(5.7)$$
Following the usual procedure, the variable occurring the least
number of times, $\phi$, is separated first, then $\theta$, then
$\beta$. leaving the final ``radial'' equation in $\rho$ for last.
The solutions of the $\rho$ equation, yielding the spectrum, depend on
the separation variable of the $\beta$ equation, a quantum number that
has no simple nonrelativistic interpretation.  The spectrum,
furthermore, for the important case $V(\rho) = -e^2 /\rho$ (which has
the nonrelativistic limit $-e^2/r$) turns out[21]
to be of the form $1/(n+{ 1\over 2})^2$, and disagrees with the 
observed results for atomic spectra.  As pointed out by Bargmann [22],
however, it is not possible to construct a representation of the
Lorentz group induced on functions with support in the full spacelike region.
There are two subregions of the spacelike measure space which are,
with translations along the $z$-axis, transitive under the subgroup
O(2,1) of the Lorentz group O(3,1), which may be described as the
interior and exterior regions obtained by constructing two planes
tangent to the light  cone at opposite sides.  The two regions are 
   described by the conditions
$$\eqalign{ x_1^2 + x_2^2 &\geq t^2  \cr
            x_1^2 + x_2^2 &\leq t^2 . \cr} \eqno(5.8)$$
Separation of variables can be carried out in both of these; the
second yields no bound states, and incorporates tunneling through the
light cone, while the first provides solutions which are bound
states, with spectrum $(1/n^2)$, consistent with experiment.  In fact
the radial equation obtained in this way coincides exactly with that of
the nonrelativistic Schr\"odinger theory, and hence every
nonrelativistic central potential problem has a corresponding
manifestly covariant form with the same spectrum [23].  The variables
are defined in this case as
$$ \eqalign{t&= \rho \sinh \beta \sin \theta \cr
            x_1 &= \rho \cosh \beta \sin \theta \cos \phi \cr
            x_2 &= \rho \cosh \beta \sin \theta \sin \phi \cr
            x_3 &= \rho \cos \theta . \cr} \eqno(5.9)$$
Separation occurs in the sequence $\phi$ first, then $\beta$ (the
separation constant provides a large degeneracy), then $\theta$, and
then $\rho$.  Hence, the radial equation contains the separation
constant which is associated with $\theta$, which carries, in this
case, the meaning of the angular momentum in the nonrelativistic
limit. These wave functions are then satisfactory to describe the
physically observable properties of a relativistic quantum system.  We
remark that, returning to $(4.8)$, we must recall that the center of
mass momentum was factored out. Writing the operator $K$ as a sum of
the two terms, 
$$ K = {P^\mu P_\mu \over 2M} + K_{rel}' , \eqno(5.10)$$ 
one can solve, in the center of mass frame, for which $ {\bf P} = 0$,
for the total energy $E$ squared. It is this quantity which is actually
measured in the laboratory. Taking the asymptotic value $-M/2$ for $K$
(established by assuming that when $V \rightarrow 0$ the
asymptotically separated particles are close to their mass shells),
and solving $(5.10)$ for $E$, one finds that the leading contribution
for small excitations is just the eigenvalues $K_{rel}'$; the next
terms constitute relativistic corrections.
\par The solutions found in our study of the bound state problem can
also be applied to the solutions of the scattering problem [24], to
the problem of calculating the selection rules and amplitudes for
radiative decay [25] and to the construction of a model for a covariant
Zeeman effect [26].  The solutions for the bound state problem in
terms of the variables $(5.9)$ are in the iirreducible representations
of O(2,1).  To achieve representations of O(3,1), the symmetry of the
differential equations, we constructed an induced representation based
on the coset space labelled by a spacelike vector $n^\mu$. We then
harmonically analyzed this induced representation over the O(3)
subgroup of O(3,1), and found that in the scattering case, one could
find the usual partial wave expansion if scattering occurred for the
vector $n^\mu$ in the beam direction.  In our study of the selection
rules, we found that the vector $n^\mu$ suffered a recoil in the
emission of electromagnetic radiation, and used this idea to
construct a covariant Zeeman interaction.
\par We have studied, so far, applications in which  strong agreement
with known results was demonstrated.  I conclude this section by
citing two crucial experiments which can distinguish a theory in which
the Einstein time is considered a dynamical variable from theories
which do not study these properties of the time. The first is the
possibility of {\it interference in time}, in exact analogy to the
Davisson-Germer experiment for spatial interference from wave
functions. Since the solutions of the equation $(4.8)$ lie in a
Hilbert space over$ R^4$, they are coherent in time as well as space
(as is well-known for electromagnetic waves);
switching the flow off and on in time, by, for example the switching
of a superconductor, has the effect of constructing a double slit in time.
It is predicted that interference phenomena could be seen for electrons
at frequencies of the order of $10^{12}$ Hz [27].  Sufficiently wide band
amplifiers at this frequency are hard to construct, but perhaps coming
into range.  A second crucial experiment, no less difficult, involves
measuring the differential cross section for scalar-scalar scattering
very close to the forward direction.  For off-shell particles, such
cross sections should be asymptotically finite, and even show
non-trivial structure (subsidiary peaks) in this region [28].  Studies
are currently under way to investigate the feasibility of carrying out
such experiments.
\bigskip
\noindent {\bf 6. Comments and Conclusions}
\smallskip
  In addition to the studies reported above, some efforts have been
made in developing a description of systems with spin.  The Hermitian
scalar product of the Hilbert space on $R^4$ precludes the use of
finite dimensional representations of O(3,1) unless they are associated
with an induced representation, of the type constructed by Wigner
[9].  However, the necessity of computing expectation values of, say,
the position $x^\mu$, acting as a derivative in momentum space, rules
out as well Wigner's use of the four-momentum to label the
coset space.  We therefore introduced a timelike four-vector [29][30] which
commutes with all observables, for this purpose, and
constructed a theory of systems with spin which resulted in a second order
equation related to Dirac's, but with completely self-adjoint
interaction between the spin and electromagnetic field.  We have
recently found a first order equation which iterates to that second
order form, and requires, to achieve this, chirally projected  wave functions
[31].
\par   The theory we have described above, which treat the time of
Einstein as a dynamical variable, provides an integrable system of
equations for the solution of problems of charges in interaction with 
electromagnetism.  Consider, for example, two charged particles
scattering through the electromagnetic force.  The usual Maxwell
formulation requires knowledge of one of the world lines in order to compute
the electromagnetic field acting on the other everywhere in spacetime.
One may then compute the trajectory of the other, and given this, the
effect of its field on the first.  The resulting motion of the first
particle may then not be consistent with the original assumption,
and the process of trial and error, or iteration, may be unstable.
The computation of the five-potential pre-Maxwell fields, however,
permits the stepwise integration of the coupled Stueckelberg and
pre-Maxwell equations.  The $\tau$ integration of the final result then 
determines the self-consistent Maxwell fields and currents [32].  Hence,
raising the dimension of the structure of the theory provides an
effective method for dealing with problems with back-reaction. 
\par  In a
similar way, the problem of constructing gravitational systems with
back-reaction, such as two stars in collision,  may be facilitated by
adding a dimension to the development of the Einstein theory.  In this
case, one generally must integrate the hydrodynamic model for the
energy momentum tensor, the right hand side of the Einstein equations,
to obtain a conserved energy-momentum tensor [15].  If we recognize
that the transformation from the locally flat coordinates of a freely
falling frame to the spacetime manifold of general relativity  may
depend on the universal world time, $\tau$, then the derivatives of
the local coordinates $\xi^\mu$ depend upon $\tau$ as well as the
(curved) $x^\mu$, i.e., 
$$ d\xi^\mu = {\partial \xi^\mu \over \partial x^\lambda} dx^\lambda +
{\partial \xi^\mu \over \partial \tau} d\tau;$$
substituting this formula into the expression for the invariant
interval leads to a {\it five}-dimensional metric tensor, somewhat
similar in form to that of Kaluza-Klein. The energy momentum tensor,
before integration over its world history, would then form the source
term for this (pre-)Einstein theory.   Some preliminary
investigations have been made of this structure [33].
\bigskip
\noindent {\bf Acknowledgements}
\smallskip
\par I wish to thank S.L. Adler for his
kind hospitality at the Institute for Advanced Study, where this
review was prepared.  
\bigskip
\noindent
{\bf References}
\smallskip
\item{1.} W. Greiner, {\it Relativistic Quantum Mechanics--Wave
Equations}, Springer Verlag, New York (1990).
\item{2.} E.C.G. Stueckelberg, Helv. Phys. Acta {\bf 14}, 372, 588
(1941); {\bf 15}, 23 (1942).
\item{3.} L.P. Horwitz and C. Piron, Helv. Phys. Acta {\bf 46}, 316
(1973).
\item{4.} T.D. Newton and E.P. Wigner, Rev. Mod. Phys. {\bf 21}, 400
(1949).
\item{5.} R. Arshansky and L.P. Horwitz, Found. Phys. {\bf 15}, 701
(1985).
\item{6.} D. Saad, L.P. Horwitz and R.I. Arshansky, Found. Phys. {\bf
19}, 1125 (1989); M.C. Land, N. Shnerb and L.P. Horwitz, J. Math.
Phys. {\bf 36}, 3263 (1995); N. Shnerb and L.P. Horwitz, Phys. Rev. A
{\bf 48}, 4068 (1993).
\item{7.} L.L. Foldy and S.A. Wouthuysen, Phys. Rev {\bf 78}, 29
(1950).
\item{8.} G.C. Hegerfeldt, Phys. Rev. D {\bf 10}, 3320 (1974); Phys.
Rev. Lett. {\bf 54}, 2395 (1980).
\item{9.} E.P. Wigner, Ann. Math. {\bf 40}, 149 (1939).
\item{10.} J. Bjorken and S. Drell, {\it Relativistic Quantum Fields},
McGraw Hill, N.Y. (1965).
\item{11.} D.J. Taylor and J.A. Wheeler, {\it Spacetime Physics}, W.H.
Freeman, San Francisco (1966).
\item{12.} I. Newton, {\it Principia}, London (1886). Andrew Motte's
translation  (1729) has been revised by F. Cajore, {\it Sir Isaac
Newton's Mathematical Principles of Natural Philosophy and his System
of the World}, University of California Press, Berkeley (1962).
\item{13.} L.D. Landau and E.M. Lifshitz, {\it Classical Theory of
Fields}, Pergamon Press, Oxford (1975).
\item{14.} M.C. Land and L.P. Horwitz, Found. Phys. {\bf 21}, 299
(1991).
\item{15.} S. Weinberg, {\it Gravitation and Cosmology}, Wiley, N.Y.
(1972).
\item{16.} The O(3,2) case treated in ref. 14.
\item{17.} R.P. Feynman, Phys. Rev {\bf 76} 769 (1949).
\item{18.} L.P. Horwitz, W.C. Schieve and C. Piron, Annals of Physics
{\bf 137}, 306 (1981).
\item{19.} J. Frastai and L.P. Horwitz, Found. Phys. {\bf 25}, 1495
(1995).
\item{20.} M. Usher and L.P. Horwitz, Found. Phys. Lett. {\bf 4}, 289
(1991).
\item{21.} J.L. Cook, Aust. Jour. Phys. {\bf 25}, 141 (1972).
\item{22.} V. Bargmann, Ann. Math. {\bf 48}, 568 (1947).
\item{23.} R. Arshansky and L.P. Howitz, J. Math. Phys. {\bf 30}, 66
(1989).
\item{24.} R. Arshansky and L.P. Horwitz, J. Math. Phys. {\bf 30}, 213
(1989).
\item{25.} M. Land, R.I. Arshansky and L.P. Horwitz, Found. Phys. {\bf
24}, 563 (1994).
\item{26.} M.C. Land and L.P. Horwitz, J. Phys. A:Math. Gen. {\bf
28}, 3289 (1995).
\item{27.} L.P. Horwitz and Y. Rabin, Lett. Nuovo Cimento {\bf 17},
501 (1976).
\item{28.} M.C. Land and L.P. Horwitz, {\it Off-Shell Quantum
Electrodynamics}, Tel Aviv University preprint TAUP-2227-95.
\item{29.} L.P. Horwitz, C. Piron and F. Reuse, Helv. Phys. Acta {\bf
48}, 546 (1975).
\item{30.} R. Arshansky and L.P. Horwitz, J. Phys. A:Math. Gen. {\bf
15}, L659 (1982).
\item{31.} B. Sarel and L.P. Horwitz, {\it A Chiral Spin Theory in the
Framework of an Invariant Evolution Parameter Formalism}, to be
published, J. Math. Phys.
\item{32.} R. Arshansky, L.P. Horwitz and Y. Lavie, Found. Phys. {\bf
15}, 701 (1985).
\item{33.} L. Burakovsky and L.P. Horwitz, {\it 5D Inflationary
Cosmology},to be published J. Gen. Rel. and Cosmology.

\vfill
\eject
\end
\bye